%% file: 000-main.tex
\documentclass[Afour,sagev,times]{sagej}

\usepackage{moreverb,url}
\usepackage[colorlinks,bookmarksopen,bookmarksnumbered,citecolor=red,urlcolor=red]{}
\usepackage{float}
\usepackage{graphicx}
\usepackage{amsmath}
\usepackage[version=4]{mhchem}
\usepackage{siunitx}
\usepackage{longtable,tabularx}
\usepackage{gensymb}
\usepackage{amsmath}
\usepackage{layouts}
\usepackage{caption}
\usepackage{subcaption}
\usepackage{amssymb}
\usepackage{nomencl}
\usepackage{xtab,booktabs}
\usepackage[section]{placeins}
\newcommand\BibTeX{{\rmfamily B\kern-.05em \textsc{i\kern-.025em b}\kern-.08em
T\kern-.1667em\lower.7ex\hbox{E}\kern-.125emX}}

\def\volumeyear{2020}

\begin{document}

\runninghead{Oruganti and Narsipur}

\title{Airfoil Lift Calculation Using Wind Tunnel Wall Pressures}

\author{Sreevishnu Oruganti\affilnum{1} and Shreyas Narsipur\affilnum{2}}

\affiliation{\affilnum{1}North Carolina State University, USA\\
\affilnum{2}North Carolina State University, USA}

\corrauth{Shreyas Narsipur, North Carolina State University,
Raleigh, NC 27695, USA.}

\email{shreya@ncsu.edu}

\input{100-abstract}

\keywords{Wind Tunnel Testing, Airfoil Lift, Wall Pressure Measurement, Chord Sensitivity Parameter, Validation}

\maketitle

\input{200-introduction}
\input{300-methodology}
\input{400-chord-sensitivity-study}
\input{500-full-validation}

\input{600-conclusions}
\input{700-acknowledgements}


\bibliographystyle{SageV}

\bibliography{trial.bib}

\section{APPENDIX}

\input{Notation}

\end{document}

%% file: 100-abstract.tex
\begin{abstract}

An experimental method to calculate lift using static pressure ports on the wind tunnel walls and its associated limits has been explored in this paper. While the wall-pressure measurement (WPM) technique for lift calculation has been implemented by other researchers, there is a lack of literature on the sensitivity of the WPM method to test section size, airfoil chord, and model thickness. Chord sensitivity studies showed that the airfoil chord plays an important role in the accuracy of the measurements and needs to be appropriately sized for a given test section dimensions for optimum performance of the WPM method. A chord sensitivity parameter ($CSP$) was formulated and a lower limit ($=0.025$) was established to relate the ideal chord-length to wind tunnel test-section dimensions to ensure best lift measuring capabilities. Finally, a combination of symmetric and cambered airfoils with thicknesses varying from $6\%-21\%$ were tested and successfully validated against reference data for a freestream chord Reynolds number range of 100,000 to 550,000. The WPM method was found to be sensitive to varying surface flow conditions and airfoil thickness and has been shown to be a viable replacement to traditional lift measurement techniques using load balances or surface pressure ports.

\end{abstract}

%% file: 200-introduction.tex
\section{Introduction}
\label{sec:introduction}
In aerodynamics, the calculation of lift is important as it greatly influences the design of aerial vehicles. Conventionally, lift on an airfoil is measured in wind tunnels using either load balances or airfoil surface pressure ports. While the load balance method is popular, the method requires constant calibration to ensure the accuracy of the force results being measured\cite{dudley1985experimental,gonzalez2011components}. On the other hand, while surface pressure measurements offer the capability of measuring the lift while giving a better insight into the surface aerodynamics, manufacturing and maintaining the models with the surface pressure ports can be very time consuming and expensive. Additionally, the pressure ports can cause surface roughness that in turn could lead to premature flow transition and affect laminar bubble formation~\cite{cole1990experimental} in steady flows and additionally affect leading-edge vortex dynamics in unsteady flow conditions, ultimately affecting the lift behavior.  

The Wall Pressure Measurement (WPM) method to measure airfoil lift has been successfully implemented in the Stuttgart University~\cite{althaus2003measurement}, NASA Langley~\cite{abbott1945summary}, and Oldenburg University wind tunnels~\cite{wolken2007dynamic,schneemann2010lift,luhur2015stochastic}. When air flows around an airfoil, gradients in the flow velocity are created around the airfoil, with a high pressure region on the lower surface of the airfoil and a low pressure region on the upper surface, causing a net upward force, called lift. Perturbations in the flow
velocity due to the presence of an airfoil in a wind tunnel test section will affect flow at the test section walls. In the WPM method, static pressures are recorded at the walls of the wind tunnel in order to capture the projected flow velocities due to the presence of the airfoil. The airfoil lift can then be measured by integrating the static pressure distributions at the walls and applying the relevant correction factors~\cite{althaus2003measurement}.

As the setup used for the WPM method is on the walls of the wind tunnel and not directly on the airfoil, as in the Surface Pressure Measurement (SPM) method, or attached to the airfoil, as in the Load Cell Measurement (LCM) method, the WPM method provides a very non-intrusive way of obtaining the airfoil lift without affecting the flow near the airfoil. 
Literature shows that the WPM method has also been effectively used to study wind tunnel blockage effects and corrections for velocity, pressure, lift and drag for vertical axis wind turbines (VAWTs), rotorcrafts, and other 2-dimensional and 3-dimensional bodies\cite{ross2011wind,garner1966subsonic,mokry1985subsonic,mokry1982subsonic,ashill1988calculation,hackett1979estimation,hensel1951rectangular,allmaras1986blockage}. The WPM method offers the additional possibility of capturing the effective shape of the airfoil as seen by the flow, as the pressure ports on the walls capture the separated or stagnant air around the airfoil along with the pressures caused by the airfoil. This can then be used to identify laminar bubbles, leading-edge separation bubbles, etc., which can be very useful for testing unsteady characteristics of airfoils~\cite{wolken2007dynamic,schneemann2010lift}. Furthermore, these wall pressures can be used in theoretical models or machine learning models to compute the airfoil surface pressures, airfoil forces and moments, and even the full flow field in the test-section.

In spite of the WPM method's popularity and possibilities, little to no literature exists on the sensitivity and predictive capabilities of the WPM method to the airfoil geometry, test section dimensions, and flow conditions. Parametric studies exploring the above conditions will help inform researchers of the testing limits and capabilities of the WPM system. In the current research, a WPM test bench is retrofitted and validated in the NCSU low-speed wind tunnel. Initial efforts were focused on conducting a chord sensitivity study to determine the optimal airfoil chord to test section dimensions ratio that will allow for the WPM system to be accurate and feasible to implement. A consistent parameter and its associated limits are explored that can relate test-section dimensions to airfoil section sizing for optimal performance of the WPM system. Subsequent efforts were focused on validating the $C_l$ measured using the WPM test bench against Surface Pressure Measurement (SPM) and Load Cell Measurement (LCM) data for eight airfoil sections with varying thickness and camber. Additionally, the prediction accuracy of the WPM method in the presence of auxiliary surfaces on the airfoil such as tripwires, vortex generators, etc., was tested.

The following section (Section~\ref{sec:methodology}) discusses the theory behind the WPM technique and its implementation at the NCSU wind tunnel facility. Sections~\ref{sec:chord_sensitivity} and \ref{sec:full-validation} present the results from the chord sensitivity and the validation studies, respectively. The final section (Section~\ref{sec:conclusion}) presents the conclusions drawn and suggests possible future directions for the current research. 

%% file: 300-methodology.tex
\section{Methodology and Experimental Setup}
\label{sec:methodology}


In this section, the methodology behind the WPM method in calculating static airfoil lift is described. Details with regard to the wind tunnel, the theory and equations used and the implemented WPM setup are presented in subsections~\ref{sec:wt}, \ref{sec:equations} and \ref{sec:wpm} respectively.

\subsection{Wind Tunnel Specifications}
\label{sec:wt}

\begin{figure}[H]
\centering
\includegraphics[width=3.25in]{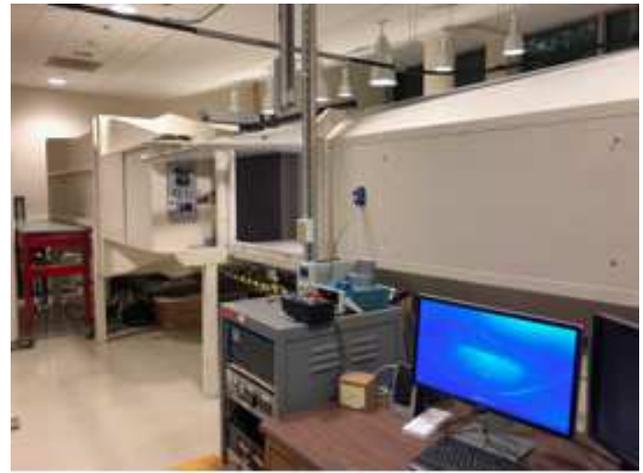}
\caption{The NC State University subsonic wind tunnel.}
\label{fig:wt}
\end{figure}

Experimental investigations were conducted in the North Carolina State University's closed-circuit, subsonic wind tunnel facility which has a 3:1 contraction ratio and a test section measuring 46~$\times$~45$\times$~32~inches (length~$\times$~width$~\times$~height). The wind tunnel is capable of reaching freestream velocities of up to 90 mph via a 3-blade, varying pitch propeller driven by a 250 horsepower electric motor. The settling chamber is equipped with a honeycomb screen and two anti-turbulence screens to ensure good flow quality. The turbulence intensity, based on turbulence sphere and hot-wire anemometry tests, has been measured to be
0.3\%. Pressure measurements were made using three, 16-port Scanivalve DSA3217 systems with a $\pm 0.05 \% $ accuracy~\cite{johnston2012investigation}. The converging and test sections of the wind tunnel is shown on Fig.~\ref{fig:wt}.

\subsection{Theory and Equations}
\label{sec:equations}


The method for deducing the airfoil lift using wall pressure ports is similar to that using surface pressure ports, with the added advantage of inexpensive airfoil models that can be rapidly manufactured. The static pressure gradient caused by the airfoil's upper and lower surfaces are measured and integrated to calculate the non-dimensionalized net upward force, or the lift coefficient, of the airfoil. However, as the circulation of the airfoil theoretically extends to infinity but the integration is performed only over the restricted length across which the pressure ports are distributed along the wind tunnel walls, correction factors as described by Althaus~\cite{althaus2003measurement} are applied to deduce the final airfoil lift coefficient. These correction factors assume that the airfoil is mounted such that centers of the airfoil and the wind tunnel test section are aligned.

\begin{figure}[t!]
\centering
\includegraphics[width=3.25in]{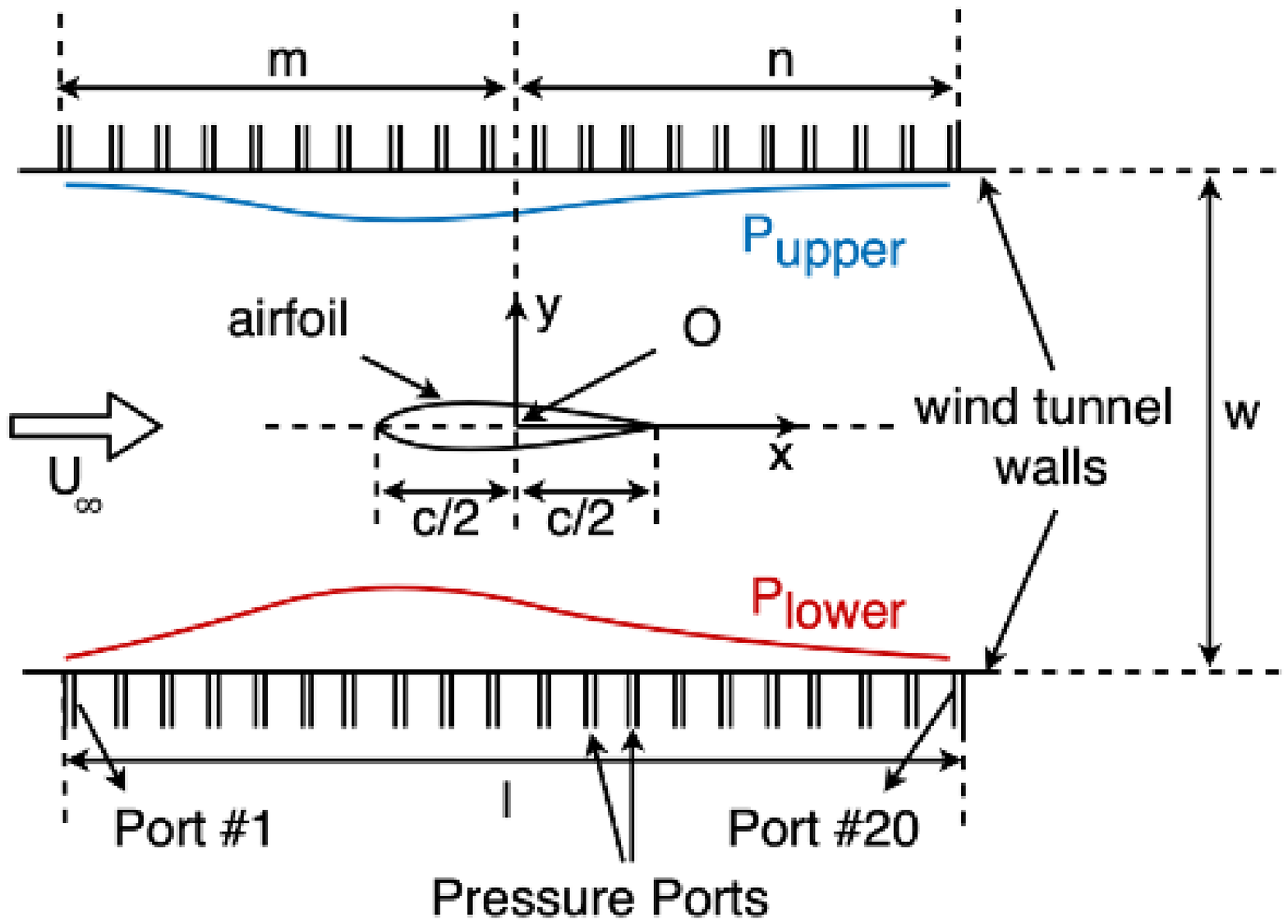}
\caption{WPM Setup Layout}
\label{fig:wpmSetup}
\end{figure}

On obtaining the static pressures from the upper ($P_{upper}$) and lower ($P_{lower}$) wind tunnel walls, parallel to the airfoil chord, the uncorrected coefficient of lift of the airfoil can be calculated as:


\begin{equation}
C_{l,wall} = \frac{1}{q_\infty}\int_m^n (P_{upper}-P_{lower})\cdot\frac{dx}{l}
\end{equation}


\noindent where $m$ and $n$ are the horizontal distances from the airfoil center to the left- and right-most wall pressure ports, respectively, and $l$ is the distance between the first and last wall pressure ports, as illustrated in Fig.~\ref{fig:wpmSetup}.

To translate the lift coefficient based on the wall pressure measurements ($C_{l,wall}$) to the actual airfoil lift coefficient ($C_{l,corr}$), two correction factors, commonly referred to as the Althaus factors\cite{althaus2003measurement}, need to be applied. These factors, determined by decomposing the airfoil's pressure distribution at the wall into a basic, constant distribution and a distribution due to the effect of the airfoil's angle of attack, are calculated using the equations,

\begin{equation}
\eta_{a} = {\int_0^c \frac{2}{\pi}\cdot\sqrt{\frac{1-\frac{x}{c}}{\frac{x}{c}}}\cdot\eta_x\cdot\frac{dx}{c}}
\label{eqn:eta_a}
\end{equation}

\begin{equation}
\eta_{b} = {\int_0^c 1\cdot\eta_x\cdot\frac{dx}{c}}
\label{eqn:eta_b}
\end{equation}

\noindent where $\eta_{x}$ is the correction factor at a given location on the wall and is dependent on the horizontal distance between the point-of-interest and the airfoil's half-chord point ($x$), width of the wind tunnel ($w$), airfoil chord ($c$), and wall pressure port distribution length ($l$):

\begin{equation}
\eta_x = \frac{2}{\pi}\cdot\arctan\bigg(\frac{e^{\frac{-\pi x}{h}}(e^{\frac{\pi n}{h}}-e^{\frac{\pi m}{h}})}{1+e^{\frac{-2\pi x}{h}}\cdot e^{\frac{\pi (m+n)}{h}}}\bigg)
\end{equation}

\begin{figure}[t!]
\centering
\includegraphics[width=.35\textwidth]{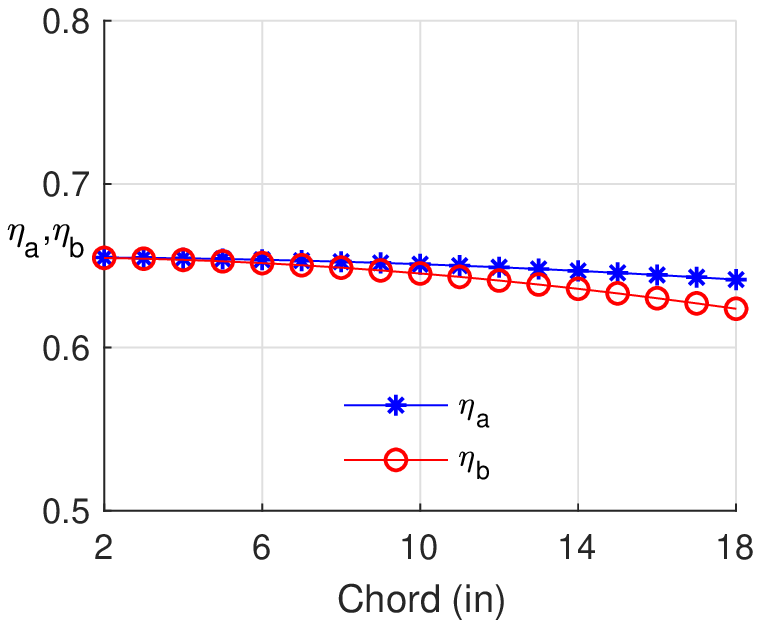}
\caption{Comparison of $\eta_a$ and $\eta_b$ for different chord lengths.}
\label{fig:eta_comp}
\end{figure}

\noindent The corrected airfoil lift coefficient can then be obtained using the equation,

\begin{equation}
C_{l,corr} = \frac{C_{l,wall}}{\eta_{a}}-C_{li}\bigg(\frac{\eta_a}{\eta_b}-1\bigg)
\label{eqn:cl_actual}
\end{equation}

\noindent where $C_{li}$ is the design lift coefficient of the airfoil. On calculating the Althaus correction factors for the current test section dimensions using Eqns.~\ref{eqn:eta_a} and \ref{eqn:eta_b}, it was observed (from Fig.~\ref{fig:eta_comp}) that $\eta_{a}\approx\eta_{b}$ for a large variation in chord lengths. Based on this information, the $(\frac{\eta_a}{\eta_b}-1)$ term in Eqn.~\ref{eqn:cl_actual} can be approximated to be zero, thereby simplifying the equation to,

\begin{equation}
C_{l,corr} = \frac{C_{l,wall}}{\eta_{a}}
\label{eqn:cl_actual_simp}
\end{equation}

\FloatBarrier

\noindent Additional wind tunnel corrections to account for turbulence and blockage effects are applied to the $C_{l,corr}$ to get the final airfoil lift coefficient. In the current experimental investigations, turbulence effects were accounted for by adjusting the freestream velocity based on the wind tunnel turbulence factor while blockage effects on the lift coefficient were corrected for using the equation described by Havelock~\cite{havelock1938lift}: 

\begin{multline}
        C_{l} = C_{l,corr}  \bigg[1-\frac{\pi c}{2w}\cdot cot\bigg(\frac{\pi d}{w}\bigg)\cdot sin(\alpha)\\ +\frac{\pi^2 c^2}{w^2}\bigg\{\bigg(\frac{2}{3}+cot^2\frac{\pi d}{w}\bigg)
    \\ + \bigg(\frac{2}{3}+3\cdot cot^2\frac{\pi d}{w}\bigg) \cdot sin^2(\alpha)\bigg\}\bigg]
\end{multline}

\noindent where $d$ is the distance between the airfoil's mid-point to the wind tunnel wall, which in the current work is $w/2$. $C_l$ is the final lift coefficient value that is obtained after applying all the necessary corrections.

\subsection{Wall Pressure Measurement Setup}
\label{sec:wpm}

\begin{figure}[H]
\centering
\includegraphics[width=.4\textwidth]{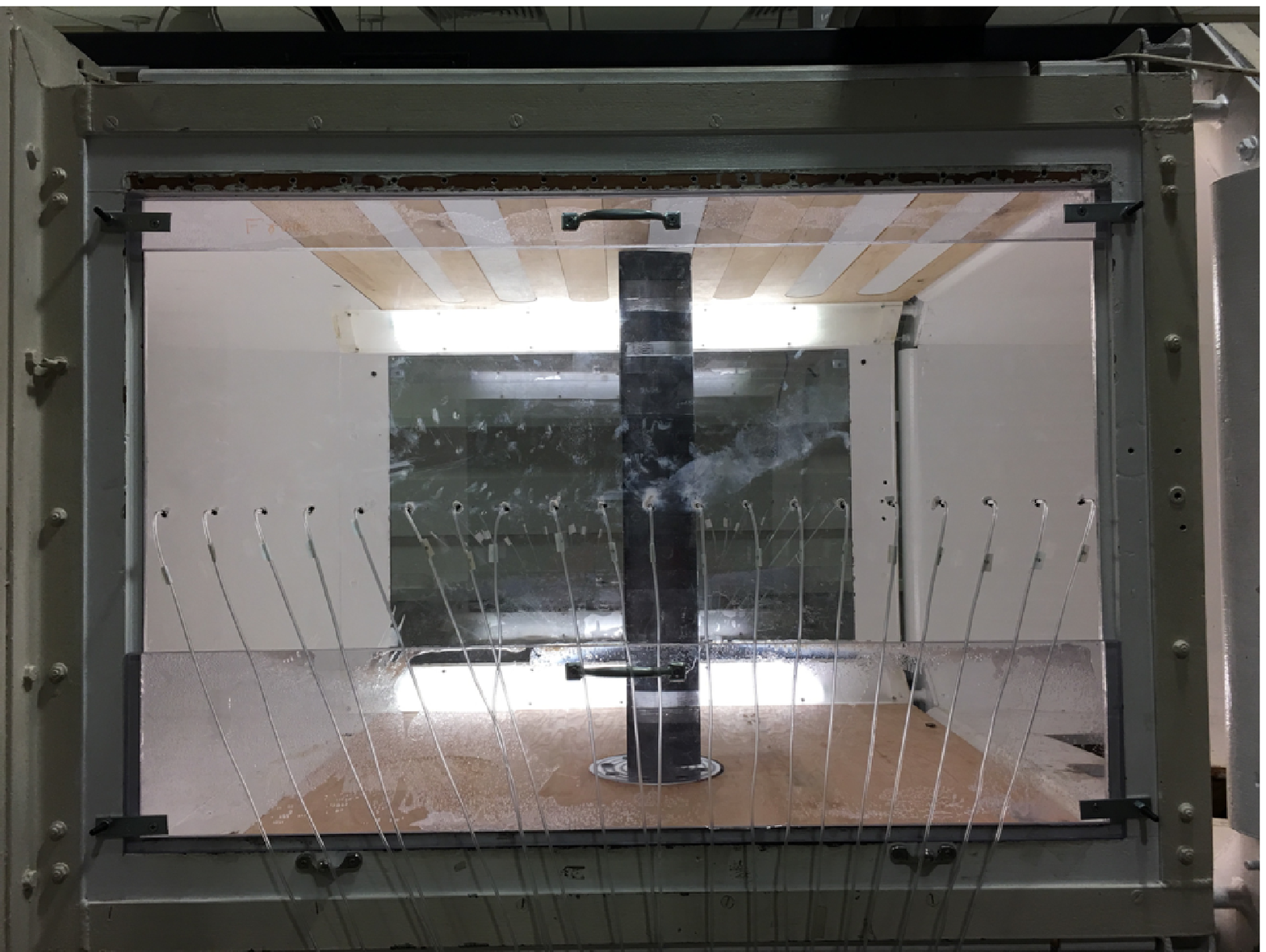}
\caption{WPM panels with the 6 inch chord LRN airfoil.}
\label{fig:setup}
\end{figure}

Guided by implementations of the WPM method in literature~\cite{althaus2003measurement,abbott1945summary,schneemann2010lift,wolken2007dynamic,luhur2015stochastic}, two plexiglass panels with 20 static pressure ports each, located at a distance of $1.95$-inches from each other along the center and spanning an overall length of $36.6$-inches, were manufactured. These panels, designed to replace the side walls of the test section, were then fitted to the wind tunnel and appropriate steps were taken to eliminate wall vibrations. Wing sections, mounted vertically on a rotating sting and spanning the entirety of the test section's height to eliminate 3D aerodynamic effects, were aligned such that the airfoil's quarter-chord point was in line with the central (10\textsuperscript{th}) wall pressure port. In order to correct for the off-set in the origin location from the half-chord to the quarter-chord point, a value of $c/4$ was subtracted from the $m$ and $n$ variables to accurately calibrate the correction factors\cite{althaus2003measurement}. All 40 pressure ports were then connected to the DSA3217 pressure scanner using flexible plastic tubing. A photograph of the WPM setup in the NCSU wind tunnel facility is presented in Fig.~\ref{fig:setup}. 

\FloatBarrier

%% file: 400-chord-sensitivity-study.tex
\section{Initial Validation and Chord Sensitivity Study}
\label{sec:chord_sensitivity}

Upon setting up the WPM test bench, initial efforts were focused on validating the setup and conducting a study to test the sensitivity of the WPM predictions to airfoil chord in order to set the test limits for the airfoil chord length ($c$) to test section length ratio ($l$), represented as a scaling ratio ($sr$), so as to guide subsequent validation studies and operation. A custom 12\% thick Low Reynolds Number (LRN) cambered airfoil was taken as the baseline geometry as a 12-inch chord version of the airfoil model with 44 surface pressure ports was available at the NCSU wind tunnel facility, thereby providing a commensurable way to compare the WPM predictions with SPM results. The symmetric, 12\% thick NACA~0012 airfoil was also considered for the initial validation and chord sensitivity study as a plethora of LCM datasets were available\cite{jacobs1937airfoil,critzos1955aerodynamic,poisson1967etude,sheldahl1981aerodynamic,michos1983aerodynamic,timmer2010aerodynamic} 
at the required Reynolds number range. Four versions of each of the two airfoils with 6-,8-,10-, and 12-inch chord lengths were 3D printed using ABS plastic with two support rods at the quarter- and three-quarter-chord locations to prevent bending or warping. The Althaus correction factors and the scaling ratios for each of the chord length's is listed in Table~\ref{tab:eta}.

\begin{table} [H]
\caption{Correction Factors and Scaling Ratios for Different Chord Lengths}
\centering
\begin{tabular}{ccc}
\hline
Chord (in.)& \parbox[t]{1.5cm}{Althaus-\\ Correction Factor\\($\eta_a$)}& \parbox[t]{1.5cm}{Scaling\\Ratio \\($sr$), \%} \\\hline
6& 0.6537 &15 \\
8& 0.6525& 20\\
10& 0.6501 & 25\\
12& 0.6491& 30\\
\hline
\label{tab:eta}
\end{tabular}
\end{table}

\begin{figure} [t!]
\centering
\includegraphics[width=.35\textwidth] {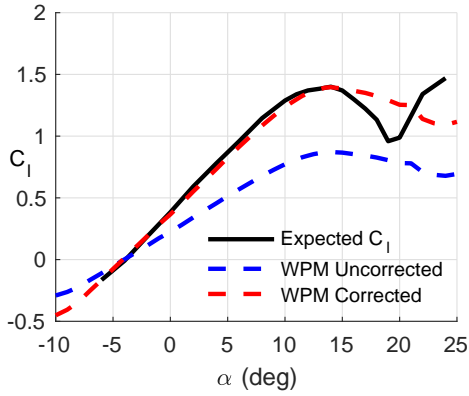}
\caption{Uncorrected and corrected WPM method results comparison with SPM results for Re = 250,000.}
\label{fig:red2-5}
\end{figure}

Figure~\ref{fig:red2-5} shows representative uncorrected and corrected lift results ($C_{l,wall}$ and $C_l$, respectively) from the WPM method for the 12-inch chord airfoil compared to the lift obtained from SPM data for the LRN airfoil at a Reynolds number of 250,000 to illustrate the steps in the WPM method to obtain airfoil $C_l$. Both the uncorrected and the corrected results follow similar trends, stall at the same angle of attack and predict zero-lift angle accurately. Additionally, corrected $C_l$ data from the WPM method is in good agreement with SPM results up to and slightly beyond stall (up to $\alpha=16$~degrees) but deviates at higher angles of attack.

\FloatBarrier

\begin{figure} [H]
\centering
\includegraphics[width=3.25in] {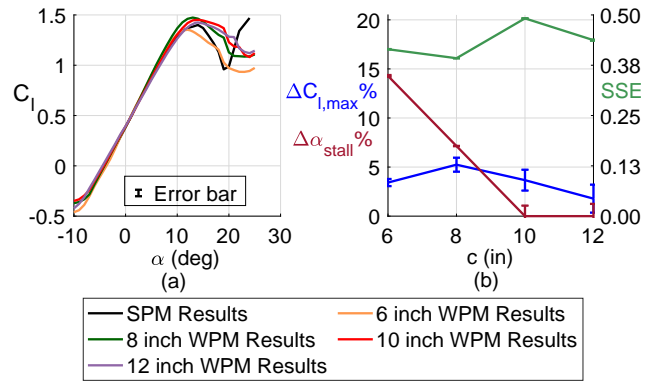}
\caption{LRN chord sensitivity study at $Re$ = 250,000. (a) $C_l$ vs $\alpha$ for 4 different chord lengths, (b) the corresponding errors in $C_{l,max}$, $\alpha_{stall}$ and SSE.
}
\label{fig:lrn_sse_2-5e5}
\end{figure}

Figure~\ref{fig:lrn_sse_2-5e5} plots the (a)~$C_l$ versus $\alpha$ results from the WPM for different airfoil chord lengths compared with SPM measurements and (b)~absolute errors in $C_{l,max}$ and $\alpha_{stall}$ along with the overall sum of square errors (SSE) against airfoil chord to investigate the WPM method's sensitivity to chord variation. The SSE between the WPM and expected results is determined using the equation,

\begin{equation}
    SSE = \sum_{1}^{N}\bigg(\frac{\sum_{1}^{N} (C_{l,exp}-C_l)}{N}-(C_{l,exp}-C_l)\bigg)^2
    \label{eqn:sse}
\end{equation}

\FloatBarrier

\noindent where $C_{l,exp}$ is the expected lift coefficient value obtained from reference data, $C_l$ is the lift coefficient from the WPM method and $N$ is the number of data points considered for taking the SSE. Observations from Fig.~\ref{fig:lrn_sse_2-5e5}(a) show that, independent of chord length, the WPM results exhibit slight to no variance in zero-lift angle of attack and the linear lift curve slope at the attached flow conditions. As $\alpha_{stall}$ is approached, the sensitivity of the WPM method is evident, with the 6-inch chord airfoil severely underpredicting $\alpha_{stall}$. 

This can be further verified from observations of Fig.~\ref{fig:lrn_sse_2-5e5}(b) where, while the error in $C_{l,max}$ predictions are $\leq5\%$ with a slight decrease for $c>8$-inches, the error in $\alpha_{stall}$ is $\approx15\%$ for the 6-inch followed by a sharp drop for the larger chord lengths and close to zero error for $c\geq10$-inches. Post-stall predictions for all airfoil chords aberrate from expected results with the magnitude of deviation captured by the SSE in Fig.~\ref{fig:lrn_sse_2-5e5}(b). Overall SSE is the least for the 8-inch chord and maximum for the 10-inch chord with the SSE predominantly depending on the post-stall lift behavior. The 6-inch airfoil accurately predicts the $C_l$ for $17\leq\alpha\leq20$~degrees while the other airfoils overpredict from right after stall to $\alpha=22$~degrees. For $\alpha>22$~degrees, all airfoils underpredict $C_l$ in comparison to SPM results. 

\FloatBarrier

\begin{figure} [H]
\centering
\includegraphics[width=3.25in] {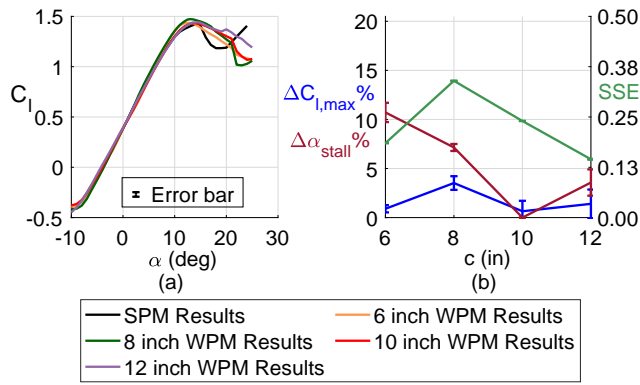}
\caption{LRN chord sensitivity study at $Re$ = 400,000. (a) $C_l$ vs $\alpha$ for 4 different chord lengths, (b) the corresponding errors in $C_{l,max}$, $\alpha_{stall}$ and SSE.}
\label{fig:lrn_sse_4e5}
\end{figure}

On performing a similar analysis at a Reynolds number of 400,000 for the LRN airfoil (Fig.~\ref{fig:lrn_sse_4e5}), we observe that while $C_l$ prediction trends for $c\geq8$-inches are similar to the $Re=250,000$ case, the predictions for the 6-inch airfoil case improve significantly. All airfoils accurately predict $C_l$ up to stall ($\alpha=14$~degrees), overpredict for $14<\alpha< 20$~degrees, and underpredict for $\alpha\geq20$~degrees. $C_{l,max}$ errors are consistently below $5\%$ for all chord lengths while $\alpha_{stall}$ errors are above $10\%$ for $c=6$-inches followed by a reduction to $\approx0\%$ at $c=10$-inches and a slight increase for the higher chord length. SSE magnitudes are lower at $Re=400,000$ with maximum (at $c=8$-inches) and minimum (at $c=12$-inches) values of 0.340 and 0.145, respectively.

\FloatBarrier

The improvement observed in $C_l$ predictions for the 6-inch airfoil is interesting and can be attributed to the low scaling ratio ($sr=15$) and high Reynolds number. At angles of attack close to stall, the airfoil experiences a loss in surface pressure due to flow separation. At lower Reynolds numbers, the magnitude of the airfoil surface pressure at these higher angles of attack are lower as compared to that at higher-$Re$ cases, due to which the resulting flow perturbation magnitudes at the wind tunnel walls are lower. Therefore, at higher Reynolds numbers, the wall pressure ports better capture the projected airfoil pressure distributions and predict the associated lift characteristics more accurately. This will not be an issue for the larger chord length airfoils as the scaling is large enough to overcome the Reynolds number dependency of the WPM method.

\begin{figure} [t!]
\centering
\includegraphics[width=3.25in] {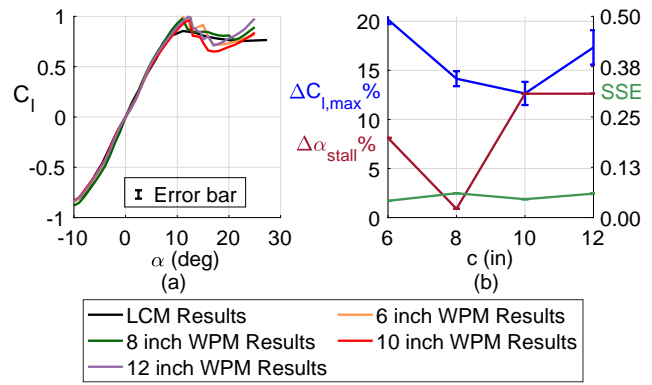}
\caption{NACA 0012 chord sensitivity study at $Re$ = 330,000\cite{jacobs1937airfoil}. (a) $C_l$ vs $\alpha$ for 4 different chord lengths, (b) the corresponding errors in $C_{l,max}$, $\alpha_{stall}$ and SSE.}
\label{fig:naca_sse_3_3e5}
\end{figure}

$C_l$ versus $\alpha$ and associated errors between the WPM predictions and LCM data for the NACA~0012 airfoil at $Re=330,000$ for different chord lengths are plotted in Fig.~\ref{fig:naca_sse_3_3e5}. While $C_l$ trends are seen to be almost independent of airfoil chord (Fig.~\ref{fig:naca_sse_3_3e5}(a)), the lift is accurately predicted until $\alpha=9$~degrees after which the WPM results slightly deviate from expected results. Additionally, the WPM results show a sharper stall behavior as compared to the LCM data. As airfoil stall is sensitive to wind tunnel conditions, the variation in stall behavior can be attributed to the difference in testing facilities. Figure~\ref{fig:naca_sse_3_3e5}(b) shows that the SSE magnitudes are comparatively lower to those observed for the LRN airfoil cases as post-stall predictions are better for the NACA~0012 airfoil. However, $C_{l,max}$ is overpredicted by at least $12\%$ ($c=10$-inches) and $\alpha_{stall}$ is $\approx1\%$ for $c=8$-inches and higher than $8\%$ for all other chord lengths.

\FloatBarrier

Based on the above results, it can be deduced that, while SSE data informs us of the overall match between the predicted and expected results, it alone is not sufficient to draw conclusions on which $sr$ combination is optimum for testing purposes, mainly evidenced by the results for the $6$-inch airfoil model which shows low SSE magnitudes but significantly under/overpredicts $C_{l,max}$ and $\alpha_{stall}$. Furthermore, though predictions for airfoils with lower $sr$ values improve as the Reynolds number increases, it is not recommended to test with similarly sized airfoil models as, in addition to their unsuitability for low Reynolds number experiments, they will require the wind tunnel to operate at higher dynamic pressure settings for a given Reynolds number, which in turn can result in high variations in testing conditions due to temperature build-up in the test section and wind tunnel vibrations. Additionally, LRN results show moderate prediction accuracy for $c=8$-inches ($sr=20$) and high prediction accuracy for $c\geq10$-inches ($sr\geq25$) while NACA~0012 results show the best accuracy for the $c=8$~inches ($sr=20$) case. Therefore, for subsequent studies in this work, airfoil test sections with 8-inch chord lengths were manufactured as it provides a good trade-off between prediction accuracy and manufacturing costs (due to low 3D printing material usage).  

\input{410-CSP}

%% file: 410-CSP.tex
\subsection{Chord Sensitivity Parameter}
\label{sec:CSP}

As one of the aims of the current work was to establish chord-sizing rules to accurately test airfoils using the WPM method, efforts were focused on defining a parameter that can be used to inform researchers on the optimal airfoil chord length for a given wind tunnel test section dimensions. From the above chord sensitivity study, it was seen that for the NCSU wind tunnel, the WPM method results were most accurate for $sr$ values above 20. However, the $sr$ value for the WPM setups at the NASA Langley~\cite{abbott1945summary} and Oldenburg University wind tunnels~\cite{wolken2007dynamic,schneemann2010lift,luhur2015stochastic} were found to be $15.4\%$ and $10\%$, respectively. Further inspection of the WPM setups in the three wind tunnels revealed that, in addition to different test section lengths, the width of the wind tunnel test sections in all three cases were different. This indicated a possible link between the wind tunnel width and accuracy of the WPM predictions. Therefore, a Chord Sensitivity Parameter ($CSP$) given by the equation,

\begin{equation}
    CSP = \frac{c^2}{w*l}
    \label{eqn:csp}
\end{equation}

\noindent was defined. Based on Eqn.~\ref{eqn:csp}, the lower limit of $CSP$ was evaluated to be 0.0389 for the current wind tunnel setup. While the limit in $CSP$ agreed with the NASA wind tunnel setup ($CSP=0.041$), it did not do so with that of the Oldenburg wind tunnel ($CSP=0.025$). It is recognized that, due to the restriction for the NCSU wind tunnel, more combinations for $c$, $w$ and $l$ could not be considered. Given this being the case, we define the lower limit of the $CSP$ to be equal to that corresponding to the Oldenburg wind tunnel and conclude that for the WPM setup, the airfoil chord should be modeled such that the $CSP\geq0.025$. While no upper limit can be established with current knowledge, there will exist one as larger airfoil-chords will have the tendency to affect wall pressures beyond the tunnel length and failing to capture them will result in incorrect $C_l$ predictions.

%% file: 500-full-validation.tex
\section{WPM Setup Validation}
\label{sec:full-validation}

In order to thoroughly validate the WPM setup, a combination of eight symmetric and cambered airfoils with thicknesses ranging from $6\%$ to $12\%$ were tested. Additionally, three airfoils were also tested in tripped-flow conditions. Subsection~\ref{sec:airfoils_tested} provides details of the airfoil geometries tested along with the associated freestream conditions. Comparison of the $C_l$ results from the WPM method with SPM and LCM data from the current research and literature are discussed in Section~\ref{sec:validation_study} to determine the validity of the current WPM setup.   

\input{510-airfoils-tested}
\input{520-validation-study}

%% file: 510-airfoils-tested.tex
\subsection{Airfoils Tested}
\label{sec:airfoils_tested}

Based on the chord sensitivity study (Section~\ref{sec:chord_sensitivity}), eight 8-inch chord wing models with different airfoil sections were manufactured with ABS plastic using 3D printing techniques and were reinforced with support rods spanning the length of the model. Table~\ref{tab:airfoils} provides details pertaining to the airfoil geometries, test conditions at which measurements were taken, and the reference data against which the WPM results were compared. Note that, for the 8-inch chord sections, the freestream Reynolds number was restricted to $550,000$ in order to ensure good flow quality. Pressure data was collected at angles of attack ranging from $-10$~degrees to $25$~degrees in $1$~degree intervals for all airfoils.   

\begin{table*}[t!]
\small\sf\centering
\caption{Airfoils Tested. 
}
\begin{tabular}{ccccc}
\toprule
Airfoil & Geometry & Thickness \% & \parbox[t]{1cm}{Reynolds \\Number} & Reference\\
\midrule
\vspace{2mm}
LRN & 
\raisebox{-\totalheight}{\includegraphics[width=0.25\textwidth, height=18mm]{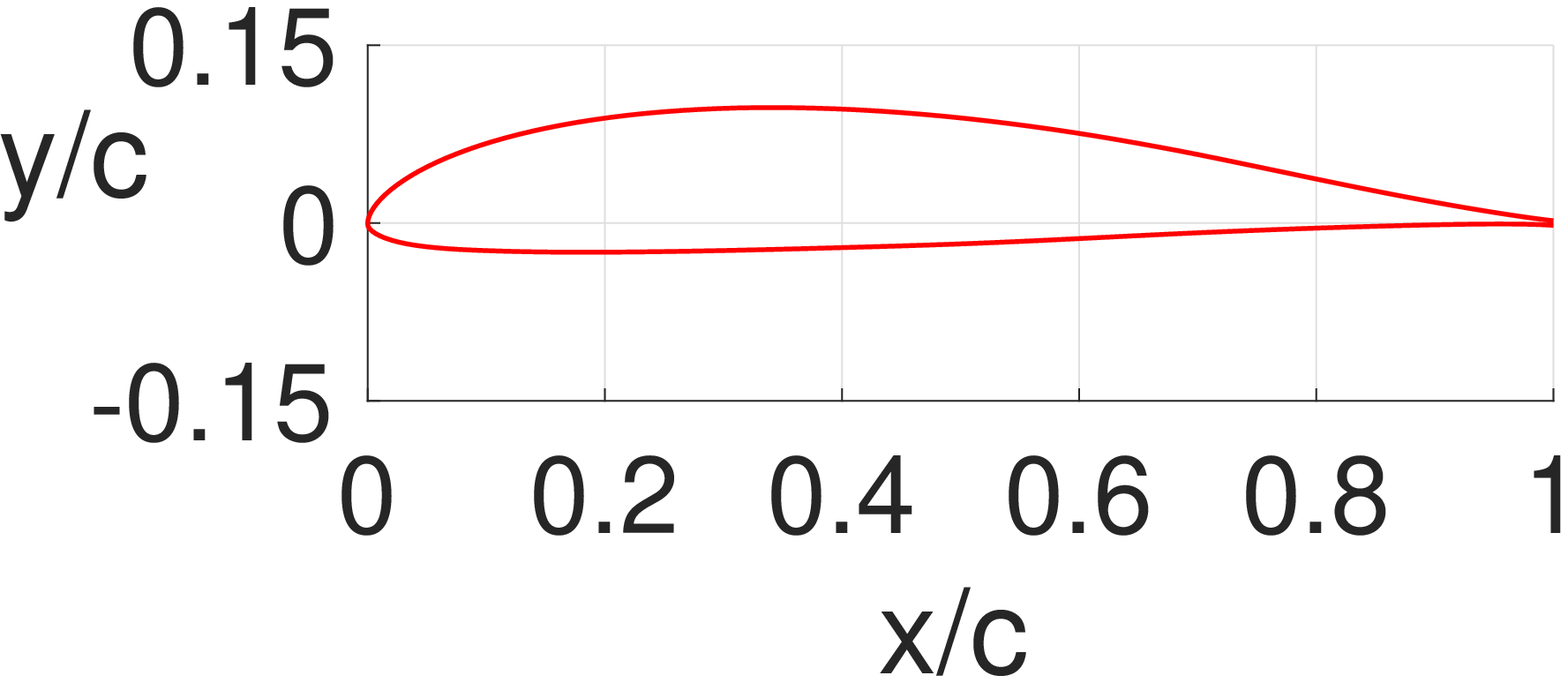}} & 12 & \parbox[t]{1cm}{250,000\\300,000\\400,000\\500,000} & SPM measurements\\
\vspace{2mm}
NACA0012 & 
\raisebox{-\totalheight}{\includegraphics[width=0.25\textwidth, height=18mm]{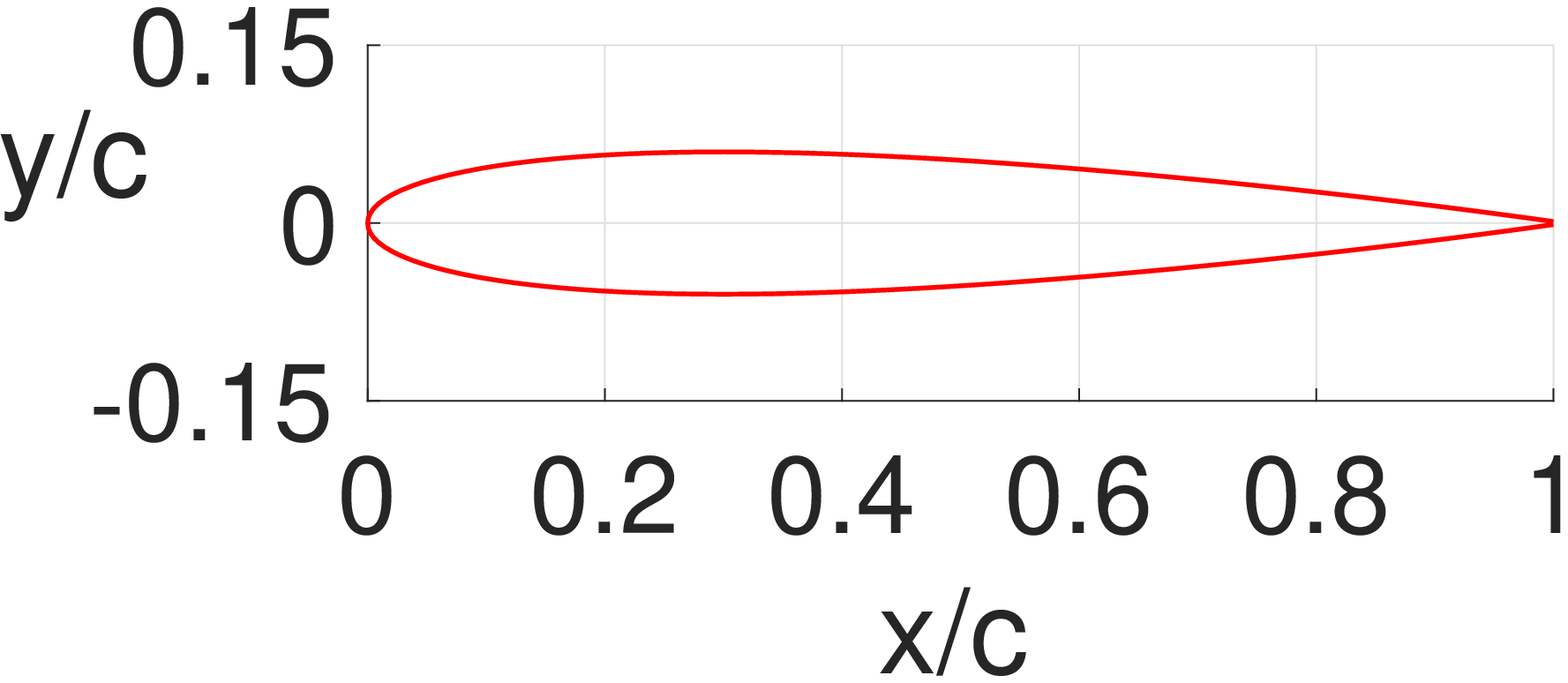}} &
12 & \parbox[t]{1cm}{170,000\\330,000\\500,000\\550,000} & \parbox[t]{2.5cm}{Jacobs et al.\cite{jacobs1937airfoil} \\ \\Critzos et al.\cite{critzos1955aerodynamic} \\ Poisson et al.\cite{poisson1967etude}}\\
\vspace{2mm}
SA7024 & 
\raisebox{-\totalheight}{\includegraphics[width=0.25\textwidth, height=18mm]{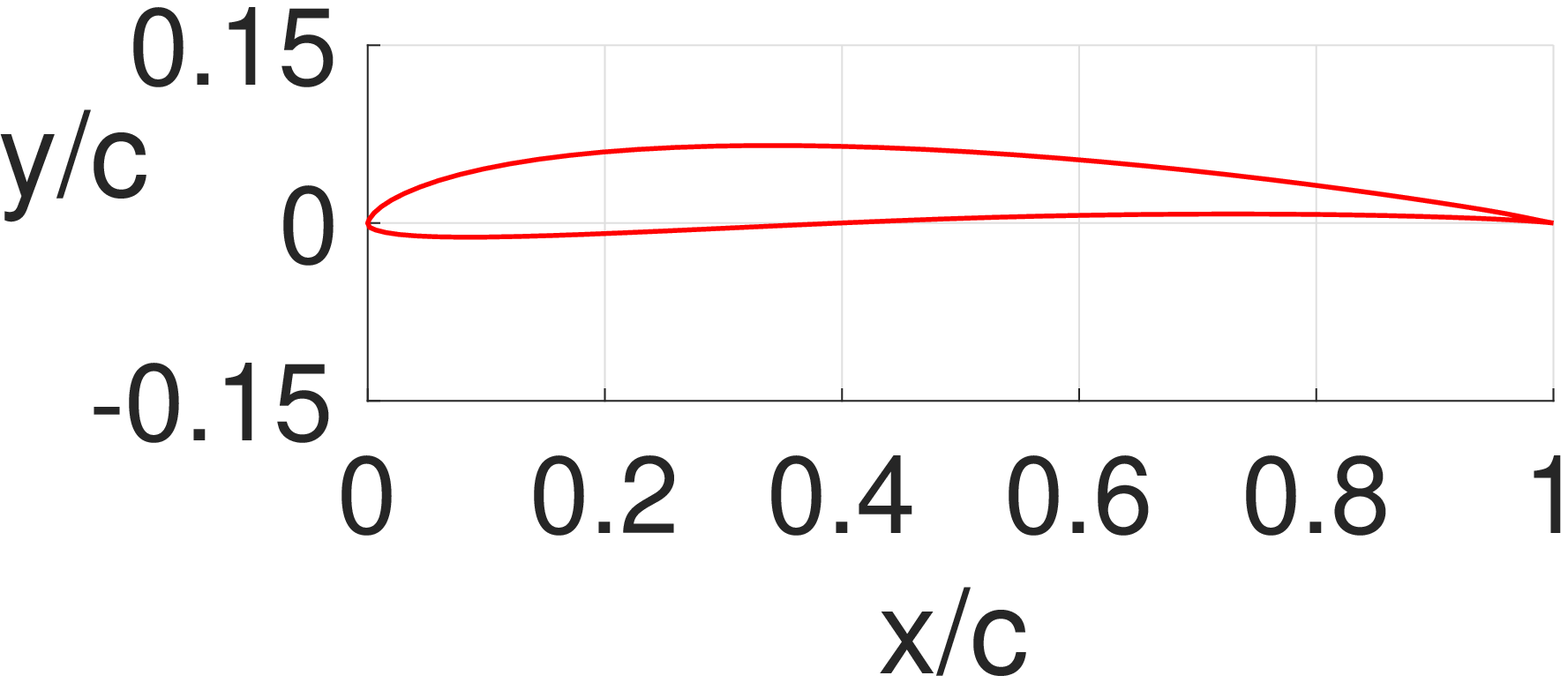}} & 
8 & \parbox[t]{1cm}{100,000\\300,000} & Gopalarathnam et al.\cite{gopalarathnam2003design}\\
\vspace{2mm}
E387 & 
\raisebox{-\totalheight}{\includegraphics[width=0.25\textwidth, height=18mm]{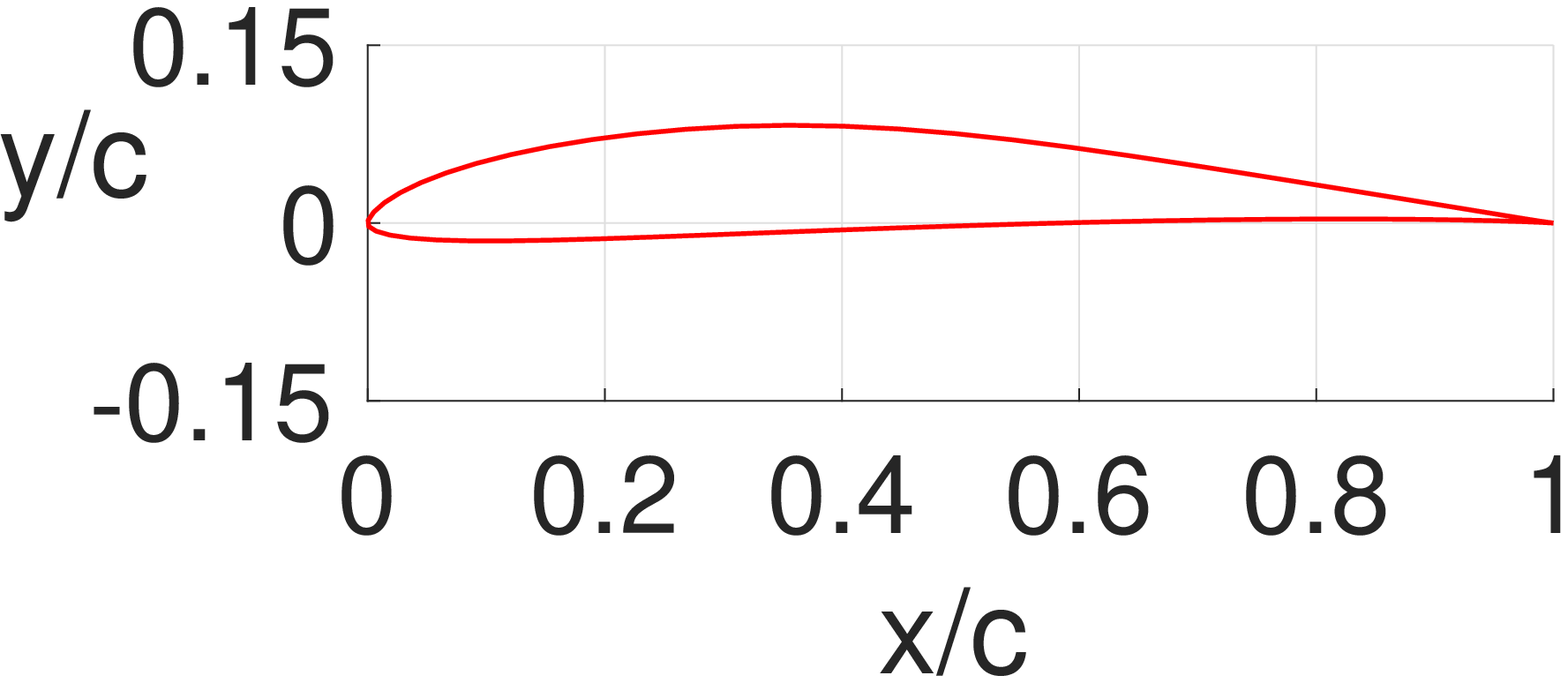}} &
9 & \parbox[t]{1cm}{200,000\\500,000} & Selig et al.\cite{seligVol4}\\
\vspace{2mm}
Clark Y & 
\raisebox{-\totalheight}{\includegraphics[width=0.25\textwidth, height=18mm]{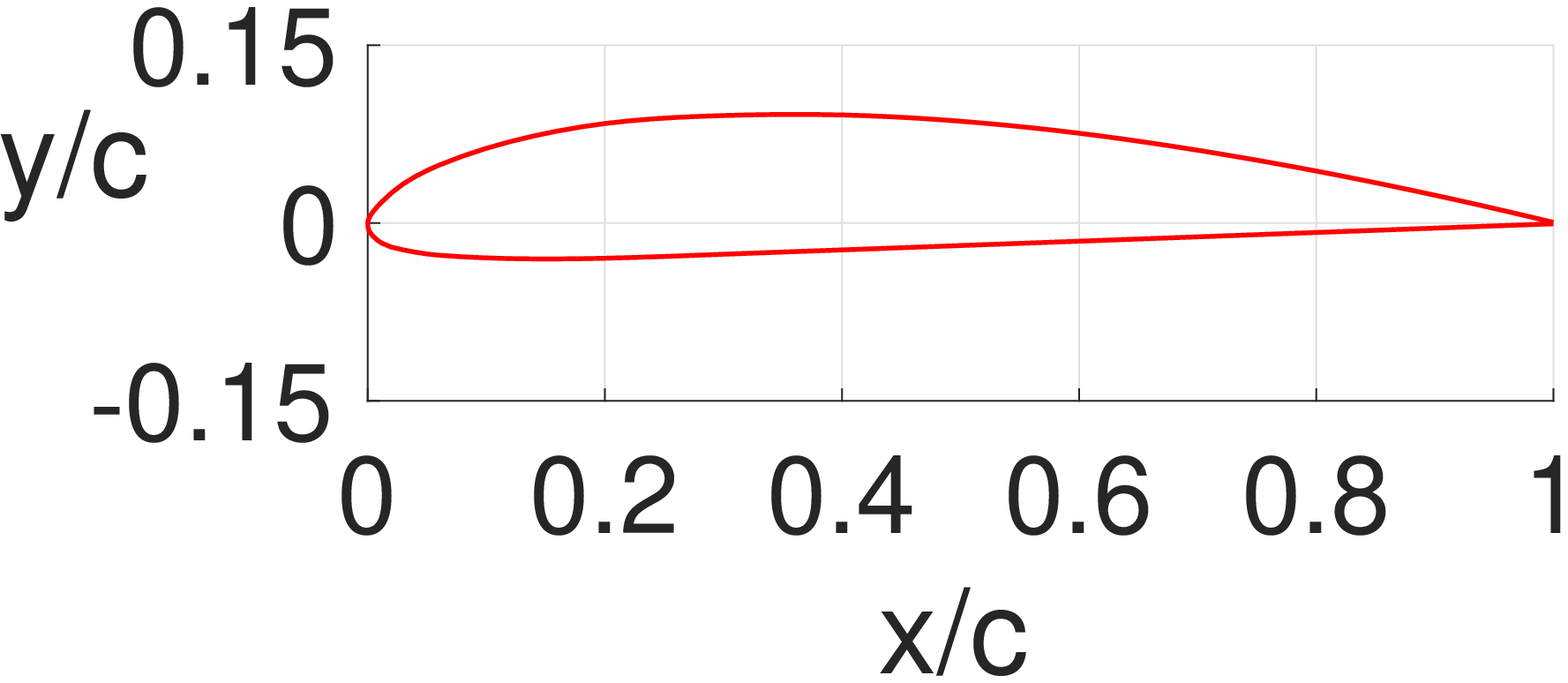}} &
12 & \parbox[t]{1cm}{300,000\\400,000} & Selig\cite{seligVol3}\\
\vspace{2mm}
Gemini & 
\raisebox{-\totalheight}{\includegraphics[width=0.25\textwidth, height=18mm]{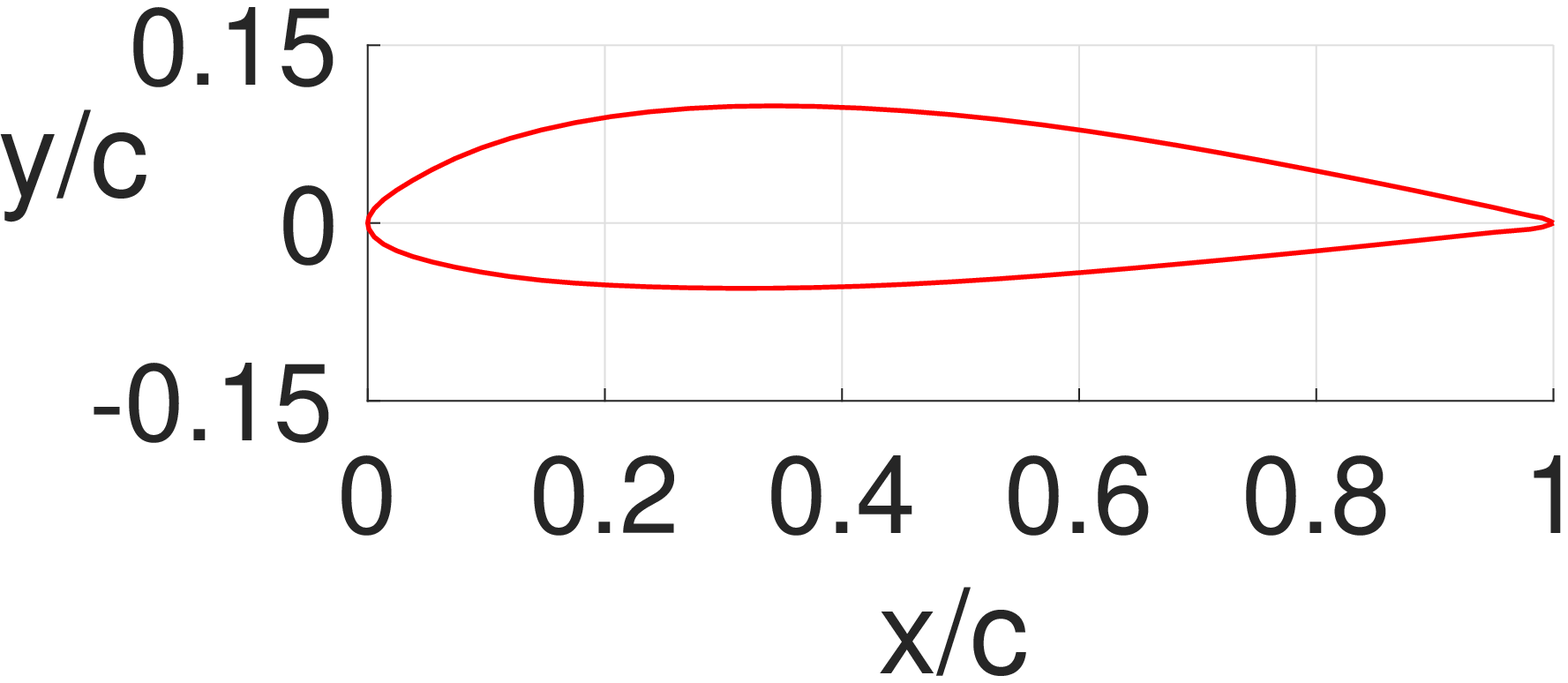}} &
15 & \parbox[t]{1cm}{200,000\\300,000} & Selig et al.\cite{seligVol1}\\
\vspace{2mm}
NACA0018 & 
\raisebox{-\totalheight}{\includegraphics[width=0.25\textwidth, height=18mm]{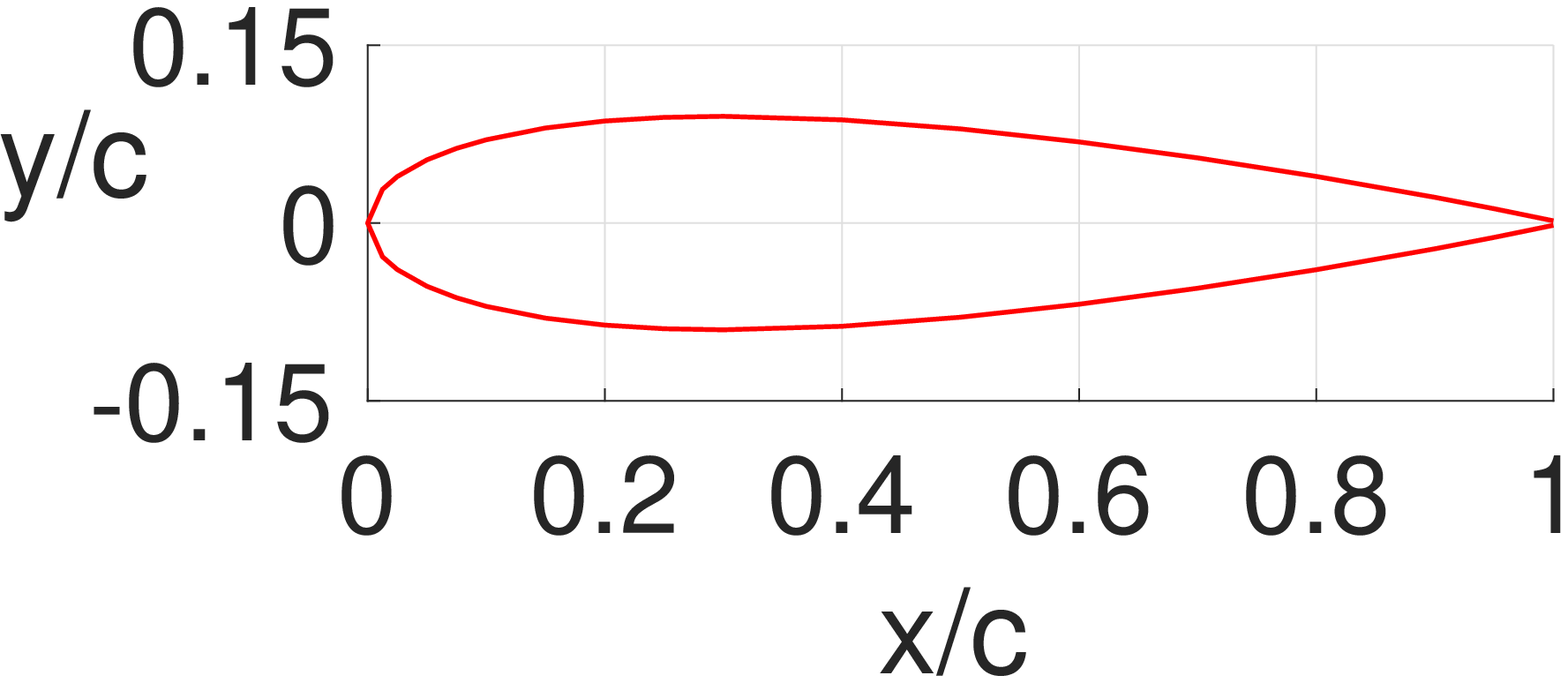}}
& 18 & \parbox[t]{1cm}{250,000\\500,000} & \parbox[t]{2.5cm}{Boutilier et al.\cite{boutilier2012parametric} \\ Timmer et al.\cite{timmer2008two} } \\
S823 & 
\raisebox{-\totalheight}{\includegraphics[width=0.25\textwidth, height=18mm]{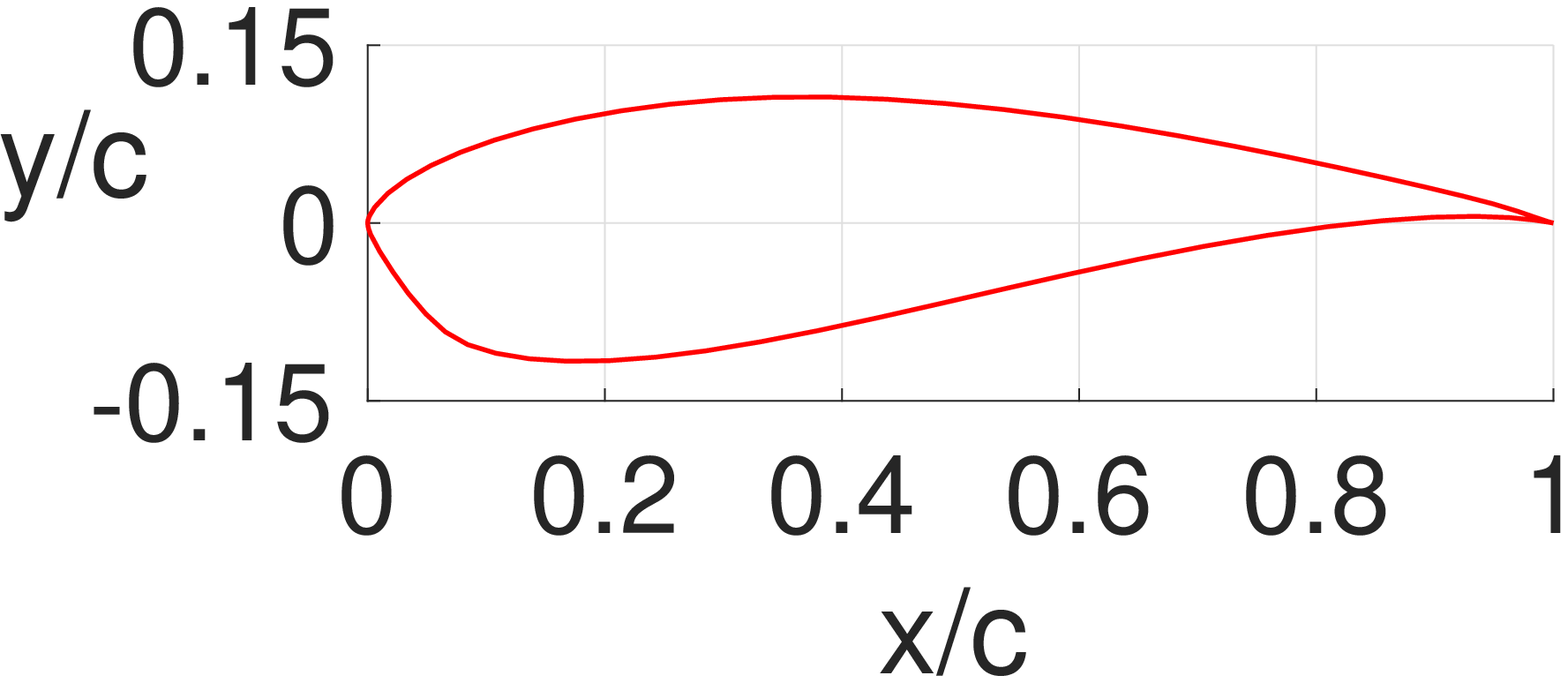}} &
21 & \parbox[t]{1cm}{300,000\\400,000} & Selig et al.\cite{seligVol1,seligVol2}\\
\bottomrule
\end{tabular}\\[10pt]
\label{tab:airfoils}
\end{table*}


%% file: 520-validation-study.tex
\subsection{Validation Study}
\label{sec:validation_study}

The current section presents the lift coefficient results predicted by the WPM test bench and validates them against experimental SPM and LCM data. For all of the validation study results shown in this section, $C_l$ vs $\alpha$ plots are shown for two or more Reynolds numbers for $\alpha$ varying from $-10$~degrees to $25$~degrees. Additionally, the error bars for the lift coefficient at every alternate angle of attack are plotted. 


\input{521-LRN}
\input{522-NACA0012}
\input{523-SA7024}

\input{524-E387}
\input{525-ClarkY}
\input{526-GEMINI}
\input{527-NACA0018}
\input{528-S823}

%% file: 521-LRN.tex
\subsubsection{LRN Airfoil ($(t/c) = 12\%$)}
\label{sec:studyLRA}

\begin{figure} [t!]
\centering
\includegraphics[width=.48\textwidth] {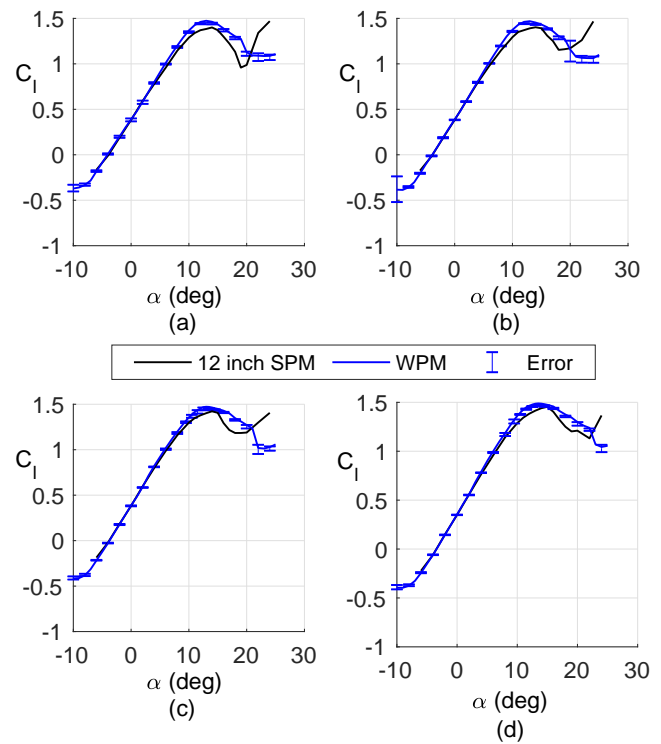}
\caption{Comparison study of SPM results vs WPM results for LRN airfoil at (a) $Re$~=~250,000, (b) $Re$~=~300,000, (c)~$Re$~=~400,000, and (d) $Re$~=~500,000\cite{johnston2012investigation}.}
\label{fig:redALL}
\end{figure}

Results of the 8-inch WPM model compared to those  obtained by integrating the surface pressure measurements of the LRN 12-inch, surface pressure model for Reynolds numbers ranging from 250,000 to 500,000 are presented in Fig.~\ref{fig:redALL}. Observations show the WPM method results match up very well with the surface pressure results at lower angles of attack ($-5$~degrees to $10$~degrees) at all Reynolds numbers. Small variations are observed for angles of attack greater than $10$~degrees with the WPM method slightly overpredicting the maximum $C_l$ at all freestream Reynolds number conditions. However, predictions are seen to improve with increase in Reynolds number. Figure~\ref{fig:redALL} also shows that key airfoil lift characteristics such as zero-lift angle, lift curve slope, and stall angle are accurately predicted for all Reynolds numbers, barring $Re=500,000$ (Fig. \ref{fig:redALL}(d)), where the WPM method predicts the stall to occur at a slightly lower angle of attack. A possible reason for the divergence could be due to the small vibrations in the model that were observed at the high Reynolds number conditions.

Post-stall $C_l$ predictions deviate from expected results, both in terms of magnitude and trends. SPM results show the $C_l$ dropping after the airfoil stalls, followed by a rapid to gradual increase, depending on the Reynolds number. However, WPM results predict a continuous drop in $C_l$ until $\alpha\approx22$~degrees after which it flattens out. The reason for this discrepancy will be investigated in future work.

%% file: 522-NACA0012.tex
\subsubsection{NACA 0012 ($(t/c) = 12\%$)}
\label{sec:study1NACA}

\begin{figure} [t!]
\centering
\includegraphics[width=3.25in] {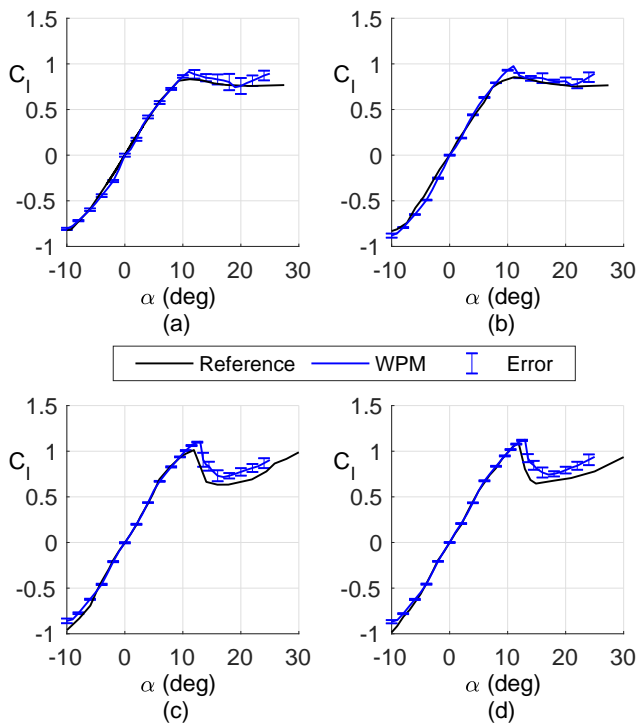}
\caption{Comparison study of LCM results vs WPM results for NACA 0012 airfoil at (a) $Re$~=~170,000~\cite{jacobs1937airfoil}, (b)~$Re$~=~330,000~\cite{jacobs1937airfoil}, (c) $Re$~=~500,000~\cite{critzos1955aerodynamic} and (d) $Re$~=~550,000~\cite{poisson1967etude}.}
\label{fig:naca}
\end{figure}

WPM results were compared with experimental $C_l$ data from LCM methods~\cite{sheldahl1981aerodynamic,jacobs1937airfoil,critzos1955aerodynamic,poisson1967etude} for the symmetric NACA 0012 for a Reynolds number range of 170,000 to 550,000, in Fig.~\ref{fig:naca}. It is observed that, at lower angles of attack ($-5$~degrees to $9$~degrees), the $C_l$ is accurately measured by the WPM method at all Reynolds numbers. For $Re=170,000$, stall angle ($\alpha_{stall}=11$~degrees) is precisely captured while $C_{l,max}$ is overpredicted by 9\% (Fig.~\ref{fig:naca}(a)). WPM results at $Re=330,000$ show that $\alpha_{stall}$ is underpredicted by $1$~degrees and $C_{l,max}$ is overpredicted by 14\% (Fig.~\ref{fig:naca}(b)). At Reynolds numbers of 500,000 and 550,000 (Figs.~\ref{fig:naca}(c) and (d)), stall angles are predicted at $14$~degrees by the WPM method, which is $2$~degrees over LCM predictions. $C_l$ is overpredicted by 7\% and 2\% for the $Re=500,000$ and $Re=550,000$ cases, respectively. Post-stall WPM results show excellent comparison with the reference data at all Reynolds number conditions, with the predictions within the expected error range of the measurements. At angles of attack below $-6$~degrees, the WPM method underpredicts the $C_l$ by a small magnitude for the Reynolds number 500,000 and 550,000 cases. Overall, WPM $C_l$ predictions are in good agreement with reference LCM experimental data for the symmetric NACA~0012 airfoil at all Reynolds number for which the validation studies were conducted. 

%% file: 523-SA7024.tex
\subsubsection{SA7024 ($(t/c) = 6\%$)}

\begin{figure} [t!]
\centering
\includegraphics[width=3.25in] {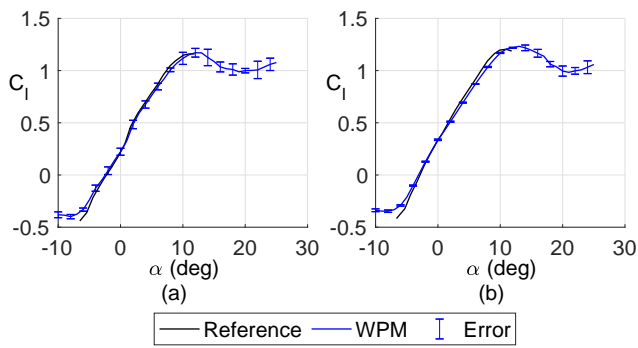}
\caption{Comparison study of LCM results vs WPM results for SA 7024 airfoil at (a)~$Re$~=~100,000 and (b)~$Re$~=~300,000\cite{gopalarathnam2003design}.}
\label{fig:sa7024}
\end{figure}

The SA7024 is a 6\% thick airfoil designed for low Reynolds number flows. Reference lift data for the SA7024 was collected in the UIUC low-speed wind tunnel and the lift measurements were made using a load balance~\cite{gopalarathnam2003design}. Figure~\ref{fig:sa7024} shows that, for all positive angles of attack, the WPM results compare well with the load balance results at both Reynolds numbers. WPM underpredicts $C_l$ for $\alpha<-5$~degrees and seems to indicate an earlier stall during negative angle of attack operation. Deviation in $C_l$, post-stall, is seen to be higher at $Re=100,000$ as compared to $Re=300,000$ and could be attributed to higher model vibrations at lower flow velocities once the airfoil has stalled. Overall, the WPM results match with reference data. Results for the current case study also show that, even though the airfoil is only 6\% thick, the flow perturbations are being accurately sensed and captured at the wind tunnel walls.


%% file: 524-E387.tex
\subsubsection{E387 ($(t/c) = 9\%$)}
\label{subsub:e387}

The Eppler 387 airfoil, commonly used in remote controlled aircraft and ultra-light powered aircraft\cite{mcghee1988experimental,mueller1985low}, was tested with and without a tripwire at Reynolds numbers of 200,000 and 500,000 and compared with LCM data from Selig and McGranahan~\cite{seligVol4}. For the tripwire cases, trips sized at 0.11\% of the chord in thickness were placed at a 2\% chord length distance from the leading edge on the upper surface and 5\% chord length distance from the leading edge on the lower surface. 

\begin{figure} [t!]
\centering
\includegraphics[width=3.25in] {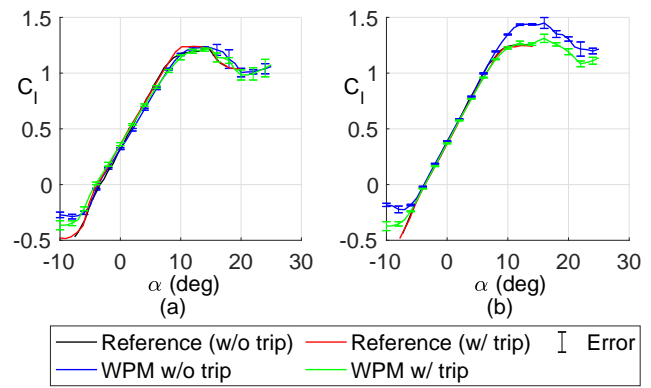}
\caption{Comparison study of LCM results vs WPM results for Eppler 387 airfoil with and without tripwire at (a)~$Re$~=~200,000 and (b)~$Re$~=~500,000\cite{seligVol4}.}
\label{fig:e387}
\end{figure}

Figure~\ref{fig:e387}(a) shows the WPM results at $Re=200,000$ accurately predicting the trends in $C_l$ for the E387 airfoil with and without tripped flow. For all positive angles of attack, while predictions for the clean flow case match very well with expected results, tripped flow results underpredict the lift for $9\leq\alpha\leq14$. $C_l$ is also slightly underpredicted at the negative angles of attack for both surface flow conditions.

\FloatBarrier

The WPM method doesn't perform so well for the clean airfoil case at $Re=500,000$ (Fig.~\ref{fig:e387}(b)), with $C_l$ being overpredicted by $\sim14\%$ for $\alpha\geq10$~degrees. Prediction performance of the WPM method for tripped flow conditions improve for positive angles of attack. However, independent of surface flow condition, the WPM underpredicts $C_l$ for $\alpha\leq-6$~degrees. This case shows that the WPM method is very sensitive to the surface flow conditions and is capable of accurately predicting the trends in $C_l$ behavior at said conditions. 

%% file: 525-ClarkY.tex
\subsubsection{Clark Y ($(t/c) = 12\%$)}
\label{subsub:clarky}

\begin{figure} [t!]
\centering
\includegraphics[width=3.25in] {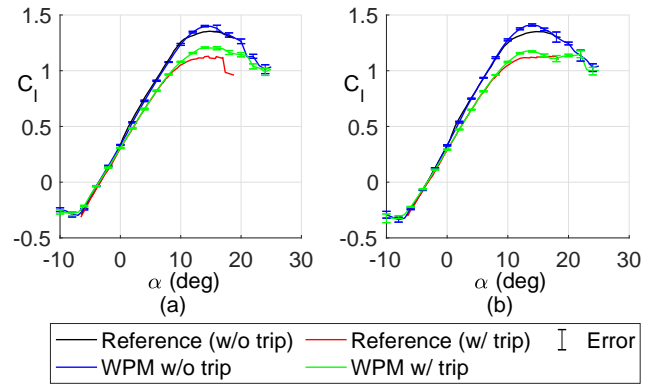}
\caption{Comparison study of LCM results vs WPM results for Clark Y airfoil with and without tripwire at (a)~$Re$~=~300,000 and (b)~$Re$~=~400,000\cite{seligVol3}.}
\label{fig:clarky}
\end{figure}

Clark Y is a flat-bottomed, 12\% thick airfoil for low-Reynolds number applications and is very commonly used for model aircrafts. The Clark Y tests were conducted at freestream Reynolds numbers of 300,000 and 400,000, with and without a tripwire. For the tripwire cases, trips sized at 0.19\% of the chord in thickness were placed at a 2\% chord length distance from the leading edge on the upper surface and 5\% chord length distance from the leading edge on the lower surface. WPM $C_l$ predictions compared with LCM results from Selig~\cite{seligVol3} are plotted in Fig.~\ref{fig:clarky}. 

\FloatBarrier

Overall, the WPM method accurately captures the $C_l$ behavior for the Clark Y airfoil, both qualitatively and quantitatively, at all tested operating conditions, with a slight deviation in the maximum lift coefficient and stall characteristics for the tripwire case at a Reynolds number of 300,000. Once again, the WPM method was successful in demonstrating its sensitivity by correctly predicting the $C_l$ behavior when the surface flow was tripped.

%% file: 526-GEMINI.tex
\subsubsection{Gemini ($(t/c) = 15\%$)}

The Gemini is a 15\% thick airfoil used in sailplanes applications and was tested at Reynolds numbers of 200,000 and 300,000. LCM data for the tests was referenced from Selig et al.\cite{seligVol1}.

\begin{figure} [t!]
\centering
\includegraphics[width=3.25in] {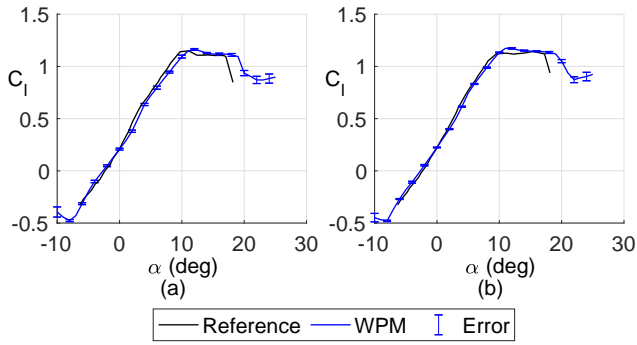}
\caption{Comparison study of LCM results vs WPM results for Gemini airfoil at (a)~$Re$~=~200,000 and (b)~$Re$~=~300,000\cite{seligVol1}.}
\label{fig:gemini}
\end{figure}

Figure~\ref{fig:gemini} shows that the WPM results are in good agreement with the expected results at both freestream Reynolds number conditions. A small deviation in the lift curve slope at $Re=200,000$, which leads to the stall angle being predicted at $11$~degrees instead of $10$~degrees, is observed in Fig.~\ref{fig:gemini}(a). On the other hand, for the $Re=300,000$ case (Fig.~\ref{fig:gemini}(b)), the lift-curve slope, stall-angle and the maximum lift coefficient are predicted accurately. At both freestream operating conditions, the sharp drop in $C_l$ post-stall is overpredicted by $3$~degrees. Overall predictions are well within the expected error range, and hence considered acceptable.

\FloatBarrier

%% file: 527-NACA0018.tex
\subsubsection{NACA 0018 ($(t/c) = 18\%$)}
\begin{figure} [t!]
\centering
\includegraphics[width=3.25in] {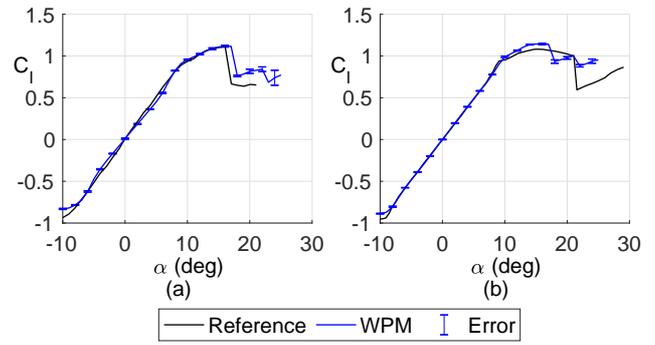}
\caption{Comparison study of LCM results vs WPM results for NACA 0018 airfoil at (a)~$Re$~=~250,000\cite{boutilier2012parametric} and (b)~$Re$~=~500,000\cite{timmer2008two}.}
\label{fig:naca0018}
\end{figure}

The symmetric, $18\%$ thick NACA~0018 airfoil was tested at Reynolds numbers of 250,000 and 500,000 and validated against results from Boutilier et. al~\cite{boutilier2012parametric} and Timmer et. al~\cite{timmer2008two}, respectively. WPM results at $Re=250,000$ (Fig.~\ref{fig:naca0018}(a)) compare well with reference data at all pre-stall angles of attack but overpredicts $\alpha_{stall}$ by $1$~degree and $C_l$ by $\sim 7\%$ for $\alpha<-6$~degrees. Post stall, the WPM predicts the $C_l$ to be $\sim 2.5\%$ above the expected results. WPM lift measurements at $Re=500,000$ ((Fig.~\ref{fig:naca0018}(b)) show excellent comparison at attached flow conditions but starts to deviate as trailing-edge boundary-layer separation increases ($\alpha\geq11$~degrees). While $C_{l,max}$ is overpredicted by $4\%$, stall angle is accurately captured. Post-stall $C_l$ trend varies with sharp drops at $18$ and $22$~degrees. Altogether, the WPM adequately captures the $C_l$ behavior with small discrepancies in the post-stall regime.

\FloatBarrier


%% file: 528-S823.tex
\subsubsection{S823 ($(t/c) = 21\%$)}

The S823 is a 21\% thick airfoil used for wind-turbine blades. For the S823, tests were conducted with and without a tripwire for Reynolds numbers of 300,000 and 400,000, and were compared with reference data from Selig et al.~\cite{seligVol1,seligVol2}. Trips, sized at 0.19\% of the chord in thickness, were placed at 2\% chord length distance from the leading edge on the upper surface and 10\% chord length distance from the leading edge on the lower surface. 

\begin{figure} [t!]
\centering
\includegraphics[width=3.25in] {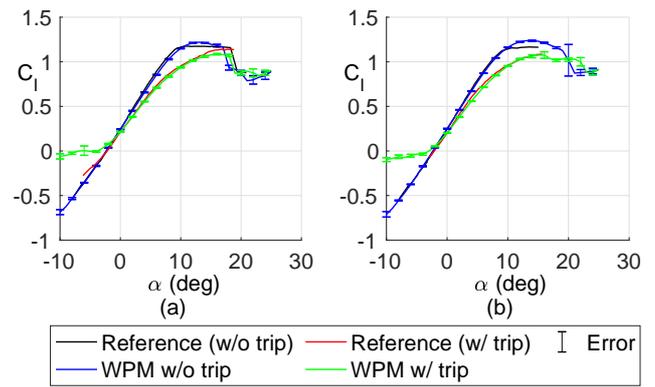}
\caption{Comparison study of LCM results vs WPM results for S823 airfoil with and without tripwire at (a)~$Re$~=~300,000 and (b)~$Re$~=~400,000\cite{seligVol1,seligVol2}.}
\label{fig:s823}
\end{figure}

From Fig.~\ref{fig:s823}, it can be observed that the WPM method results compare well with the expected results at all angles of attack even after stall for the cases without tripwire. With tripped flow, the results agree well for all positive angles of attack but deviate for negative angles of attack, with WPM results showing earlier flow separation ($\alpha\leq-1$~degree) and overpredicting $C_l$ by $\sim 15\%$. Post-stall $C_l$ at the negative angles of attack for the tripped flow case stays almost constant. Overall, the WPM results comparing well with reference data shows the effectiveness of the method in testing very thick airfoil sections.

\FloatBarrier

%% file: 600-conclusions.tex
\section{Conclusion}
\label{sec:conclusion}

This research paper discussed the implementation of the Wall Pressure Measurement (WPM) technique to calculate lift as a feasible and quick method to experimentally measure airfoil lift behavior. Though the WPM technique has been implemented in other wind tunnels, little literature exists with regard to the sensitivity of the method to variation in airfoil chord length and thickness. Initial efforts were focused on conducting a chord sensitivity study to get the optimum airfoil chord to wind tunnel test section length ratio for effective performance of the WPM method. This was followed by a detailed validation study at freestream Reynolds number conditions ranging from 100,000 to 550,000, to test the effectiveness of the WPM test bench in predicting lift data for various airfoil geometries with varying maximum thicknesses and camber.

Results from the chord sensitivity study showed that the WPM method is highly sensitive to the chord of the airfoil and the best results were obtained in the NCSU wind tunnel when the airfoil chord was above 20\% of the test section length. However, the $20\%$ scaling ratio did not agree with other implementations of the WPM method and indicated that merely the airfoil chord and test section length were not enough to effectively represent the sensitivity of the WPM method to test section and model geometry. A new design parameter called $CSP$ was formulated, taking into account the width of the test section, and was successful in setting a lower limit on the ideal chord-length of the airfoil that would produce the best WPM results for any test-section dimensions. Based on current work and literature, the lower limit of $CSP$ was found to be 0.025 which provides a good starting point to model the WPM setup for optimum performance.

Validation study results showed that the $C_l$ predicted by the WPM method is comparable to those measured using the SPM and LCM methods for airfoil geometries with thicknesses ranging from $6\%$ to $21\%$, with the added advantage of testing with a cheaper and simpler setup that can capture all aerodynamic features without introducing any surface roughness, flow intrusions and arduous calibration procedures. WPM results showed good comparison for cases with tripped flow, implying that even minute flow perturbations were accurately being captured by the method. In a few cases, an increase in Reynolds number resulted in smaller deviations in the results from expected values. In conclusion, the WPM method can serve as a viable, non-intrusive, and inexpensive replacement to existing lift measurement techniques.

Future iterations of this research work will focus on setting an upper limit to the $CSP$ with possible extensions focusing on backing the airfoil surface $C_p$ from the wall pressures using theoretical panel methods and machine learning approaches. Implementing the WPM technique for unsteady lift measurements and obtaining the effective shape of the airfoil in unsteady flows can also be explored.

%% file: 700-acknowledgements.tex
\section{Acknowledgment}
\label{sec:acknowledgment}
The authors wish to thank Jim Dean from the College of Design, NC State University, for manufacturing all the required setup, Shaphan Jernigan from the MAE Dept., NC State University, for 3D printing the required airfoils, Samuel Richardson for helping with manufacturing, and the Experimental Aerodynamics team for helping with wind tunnel testing.

%% file: Notation.tex
\textbf{Notation}


\vspace{2mm}

\begin{xtabular}{ll}
$\alpha$ & angle of attack \\
$\alpha_{stall}$ & stall angle of attack \\
$b$  & blockage of airfoil, chord/height of test-section ($c/w$) \\
$c$  & chord of airfoil \\
$C_l$ & coefficient of lift of airfoil\\
$C_{l,max}$ & maximum coefficient of lift of airfoil\\
$C_{li}$ & design coefficient of lift of airfoil\\
$C_p$ &    coefficient of pressure \\
$CSP$ &    chord sensitivity parameter \\
$\eta_a$ & Althaus correction factor for angle of attack \\
$\eta_b$ & \parbox[t]{5in}{ Althaus correction factor for constant pressure distribution} \\
$\eta$ & Althaus correction factor \\
$\eta_x$ & correction factor for a point vortex \\
$l$ & distance fromm first pressure port to last pressure port \\
$m$ & distance from center of airfoil to left most pressure port  \\
$n$ & distance from center of airfoil to right most pressure port \\
$\Delta P$ & change in pressure \\
$P_{upper}$ & static pressure of upper surface \\
$P_{lower}$ & static pressure of upper surface \\
$P_R$ & resulting pressure coefficient for tunnel walls\\
$q_\infty$ & freestream dynamic pressure, $\frac{1}{2} \rho V_\infty^2$\\
$Re$ & chord Reynolds number\\
$\rho$ & air density\\
$sr$ & scaling facratiotor, $\frac{c}{l}*100$\\
$l$ & test section length\\
$V$ & freestream velocity\\
$w$ & test section width \\
\end{xtabular}%